# Mapping forest age using National Forest Inventory, airborne laser scanning, and Sentinel-2 data

Johannes Schumacher, Marius Hauglin, Rasmus Astrup, Johannes Breidenbach

Norwegian Institute of Bioeconomy Research, National Forest Inventory

## Abstract

The age of forest stands is critical information for many aspects of forest management and conservation but area-wide information about forest stand age often does not exist. In this study, we developed regression models for large-scale area-wide prediction of age in Norwegian forests. For model development we used more than 4800 plots of the Norwegian National Forest Inventory (NFI) distributed over Norway between 58° and 65° northern latitude in a 181,773 km$^2$ study area. Predictor variables were based on airborne laser scanning (ALS), Sentinel-2, and existing public map data. We performed model validation on an independent data set consisting of 63 spruce stands with known age.

The best modelling strategy was to fit independent linear regression models to each observed site index (SI) level and using a SI prediction map in the application of the models. The most important predictor variable was an upper percentile of the ALS heights, and root-mean-squared-errors (RMSE) ranged between 3 and 31 years (6% to 26%) for SI-specific models, and 21 years (25%) on average. Mean deviance (MD) ranged between -1 and 3 years. The models improved with increasing SI and the RMSE were largest for low SI stands older than 100 years. Using a mapped SI, which is required for practical applications, RMSE and MD on plot-level ranged from 19 to 56 years (29% to 53%), and 5 to 37 years (5% to 31%), respectively. For the validation stands, the RMSE and MD were 12 (22%) and 2 years (3%).

Tree height estimated from airborne laser scanning and predicted site index were the most important variables in the models describing age. Overall, we obtained good results, especially for stands with high SI, that could be considered for practical applications but see considerable potential for improvements, if better SI maps were available.

## 1. Introduction

Forest stand age is a key parameter in designing both forest management and forest conservation strategies. Determining forest age is not a trivial task and can be difficult in complex forest structure but simpler in more even-aged forest. A description and characterization of the different types of forest age can be found in Chirici et al. (2011). Forest age can be determined based on stem cores taken at individual reference trees (Grissino-Mayer 2003). Counting the number of year rings in the stem core extracted close to the ground leads to a good estimate of tree age. However, this is a tedious and time-consuming method. Alternative technologies are required to estimate forest stand age for larger areas to be of practical use for forest management and conservation strategies. For example, in Norway the age of stands is often unknown, there are no public maps with reliable estimates of forest age, and even in areas with forest management inventories age is often one of the most uncertain parameters.

Growing processes over time result in specific tree dimensions and forest structure and are determined by a combination of historic management and abiotic factors. Management include tree species, genetic material, establishment history, stand density, and past harvesting activities. Abiotic factors are environmental conditions including topography, soil type, and macro and micro climatic variables. These characteristics determine the growth and production potential of a site for a given tree species and result in trees of greatly varying dimensions given the same age. Site index is typically used to describe this productivity and is tree-species specific. Site index is commonly determined by measuring the height and age of dominant trees in even-aged stands found on that site (Skovsgaard and Vanclay 2008). Other methods for estimating site index make use of climate (Nothdurft et al. 2012; Sharma et al. 2012) or remotely sensed data (Socha et al. 2017; Kandare et al. 2017).

Forest stand age and site characteristics described by site index determine the status of forest height, structure and density. Therefore, stand height and structure together with site characteristics can be used to estimate stand age. Proxies for forest stand height and structure can be estimated from remotely sensed data, such as airborne laser scanning (ALS) (Næsset 1997; Nord-Larsen and Riis-Nielsen 2010; Hudak et al. 2014; Mura et al. 2015; Guo et al. 2017) or optical data (Kayitakire et al. 2006; Mora et al. 2013; Lang et al. 2019), which often exist for large areas. In Norway, the site characteristics climate, and water and light availability are largely related to geographical location, height above sea level (asl), distance from coast, and terrain slope, which determine temperature, precipitation, water and nutrient availability, and length of growing season (Antón-Fernández et al. 2016).

Previous research demonstrated that forest stand age can be modelled using spectral data (Jensen et al. 1999; Reese et al. 2003; Buddenbaum et al. 2005; Kayitakire et al. 2006; Dye et al. 2012), 3D data from ALS (Maltamo et al. 2009; Racine et al. 2014; Zhang et al. 2014), and a combination of both (Straub and Koch 2011). Kayitakire et al. (2006) used image texture from IKONOS-2 satellite images with 1 x 1 m ground resolution and linear modelling to explain forest stand age and other structure-related variables. Their study site was in Belgium and included 29 sample plots in even aged Norway spruce stands with age between 27 and 110 years. Age was best explained by the correlation texture calculated within a 15 x 15 m window, resulting in $R^2$ of 0.81 with a mean absolute error of 10 years. Racine et al. (2014) used ALS data and the k-nearest neighbour (kNN) approach to estimate forest stand age based on 158 forest plots in managed boreal forest in central Quebec, Canada. The mean plot age ranged from 11 to 94 years. The best model combining ALS based forest structure variables and ALS based variables describing site characteristics resulted in $R^2$ of 0.83 and root-mean-squared-error (RMSE) of 8.8 years. Maltamo et al. (2009) used ALS data and the k-most similar neighbour approach on 335 NFI plots in a 22,000 ha study site in Finland and reported RMSE of 23.5, 18.8, and 18.7 years for age of pine, spruce, and deciduous plots, respectively. Straub & Koch (2011) used both airborne ALS and multispectral variables to model forest stand age in a small study area (9.24 km$^2$, 108 forest stands, 300 inventory plots) in south-west Germany using linear regression, reporting

RMSE and RMSE% of 19.7 years and 28.8%, respectively. Buddenbaum et al. (2005) used hyperspectral HyMap data from the spectral angle mapper to classify Norway spruce and Douglas fir age classes on a located within one HyMap scene covering 2.5 km x 10 km in south-west Germany. Their best result of 81% overall accuracy was achieved with the maximum likelihood classifier for the four age categories 10-30, 30-50, 50-80, and >80 years. Reese et al. (2003) performed estimation of forest stand age using Landsat 7 satellite data, field data from the Swedish NFI, and the kNN approach in south-western Sweden. In this area, field data for 89 Norway spruce dominated stands ranging from 6 to 106 years were available. RMSE of predicted age for these stands was 12 years or 23%. Maltamo et al. (2019, pers. communication) found that modelling age of stands using ALS for forests older than 100 years was infeasible. For forest stands below 100 years, their models resulted in RMSE of 9 and 10 years when using ALS data and site index, and ALS data alone, respectively.

In other studies outside the temperate and boreal zone, Dye et al. (2012) used spectral and texture information from high resolution satellite QuickBird images and random forest to predict the age of pine forests in the west of South Afrika. They used 142 sample stands, and age ranged from 4 to 24 years. Their normalized out-of-bag errors ranged from 28% to 34%. Jensen et al. (1999) modelled coniferous forest age of a small study site in Brazil using Landsat TM satellite data and regression and artificial neural network approaches. Main tree species was loblolly pine and tree age ranged from 1 to 40 years. Percentage of stands with absolute age errors below 2 years were up to 83% of all stands for multiple regression modelling and up to 98% of all stands for the best artificial neural network. In a study in central Italy comprising 128.402 ha forest in various growing conditions, Frate et al. (2015) used multispectral satellite imagery and 304 field plots to first model timber volume using the kNN approach, and subsequently, used inverted yield models to predict forest age. On 305 independent validation stands covering 3,137 ha and stand age from 1 to 127 years with mean of 52 years, they obtained forest age estimates with RMSE of 16 years (30%). Zhang et al. (2014) used forest height from spaceborne ALS and non-linear regression modelling to predict forest age in China at 1 km resolution. The authors fitted first biomass-height and then age-biomass models using field observations from 3543 forest plots, and reported $R^2$ of 0.6 and 0.7, respectively. However, the spatial resolution was 1 km, which is too coarse for operational forest management. To the best of our knowledge there are no large-scale studies predicting forest stand age that can be used for practical forest management.

The objectives of the present study were to (i) model and map forest stand age of Norway spruce (*Picea abies* (L.) H.Karst.), Scots pine (*Pinus sylvestris* L.), and broadleaved (mostly downy birch (*Betula pubescens* Ehrh.)) dominated forest using National Forest Inventory (NFI) sample plots with ALS and optical satellite variables; and (ii) to validate the result on an independent data set consisting of forest stands with known age. Even though we conducted all analyses for all three tree species, we will focus on spruce here and present the results for pine and birch in the appendix. We believe this to be the first attempt to model and predict forest age at a regional scale using national laser scanning campaigns.

## 2. Material and Methods
### 2.1 Study area
The 181,773 km$^2$ study area is located in Norway between 58° and 65.3° northern latitude and was determined by the availability of airborne ALS coverage (Figure 1). Growing conditions vary considerably with latitude and elevation. The natural tree line is at around 1100 m above sea level (asl) in southern Norway and around 130 m asl in the north. Depending on these factors, climate zones range roughly from subarctic in the north and east, oceanic at the coast, and continental in the south-east. Tree species of main economic interest are spruce and pine, making up for the majority of biomass and timber volume. Birch-dominated forests are most abundant in terms of forest area and mainly occur as an early succession species after harvests or in high elevation forests.

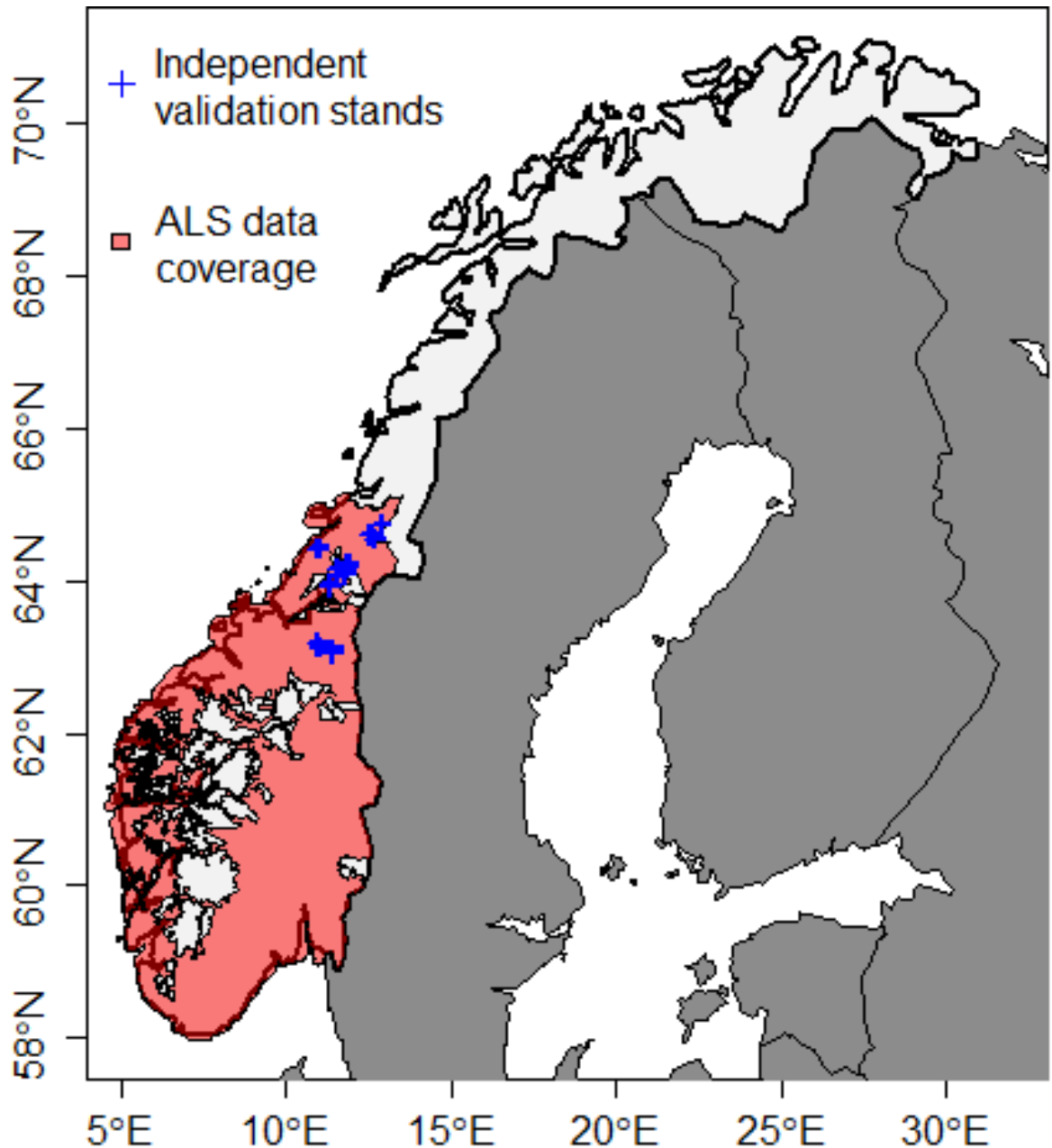

*Figure 1: Map of the study site in Norway; ALS data coverage is displayed in red, and location of independent validation stands as blue crosses.*

## 2.2 Data

### 2.2.1 National Forest Inventory data

We used the permanent sample plots of the Norwegian NFI as reference data (Tomter et al. 2010). In the study area, the NFI is based on a systematic grid of 3 km x 3 km in the low-land region and 3 km x 9 km in the low-productive, birch-dominated mountain region. For trees with a diameter at breast height >= 5 cm (dbh, 1.3 m above ground), parameters are measured on circular plots with a size of 250 m². Tree heights of approximately 10 trees are measured using a Vertex hypsometer (Haglöf 2007). The heights of the remaining trees are predicted using locally calibrated height-diameter functions, and arithmetic mean height for each plot is calculated.

Stand parameters such as age and site index are determined on circular sample plots of 1000 m². Each plot center is permanently marked with a metal pole buried in the ground with known coordinates determined by a global navigation satellite system (GNSS) device. The Norwegian NFI completes one full cycle every five years, i.e. each year one fifth of all plots are visited and forest variables measured. Relevant variables for the present study are stand age, mean height, and site index (SI).

Stand age is determined for each plot at one or more representative trees just outside the 250 m²-plot boundary by taking one or more stem cores using an increment corer. The biological age, rather than chronological age, is recorded that corrects years of suppression below canopy after germination. Alternatively, the number of whirls is counted in young forest where this is possible. In forests that consist of either one or more than two layers, age is the basal-area weighted age of all trees. In two-layered forests, age is the basal-area weighted age of all trees in the dominant layer.

SI is determined in classes of of 6, 8, 11, 14, 17, 20, 23, and 26 which describe the height (m) of the top 100 trees per ha at age 40. To this end, height and age of a tree representative of the 10 largest trees on the 1000 m² plot is measured and SI is determined based on SI curves (Tveite 1977).

We used NFI plots located in stands dominated by spruce, pine, and broadleaved species (defined as plots with >= 75% timber volume of each tree species, respectively). The major tree species in broadleaved dominated forests is downy birch (*Betula pubescens* Ehrh.), and in the following we only refer to birch when addressing broadleaved dominated forest. From these plots, we only selected NFI plots in productive forest (yearly volume increment > 0.1 m³ / ha), and removed plots with canopy emergent seed-trees, resulting in 2121 spruce, 1779 pine, and 929 birch dominated plots that were used for modelling. Age for these plots ranged from 3 to 270, 4 to 287, and 3 to 223 years, respectively (Table 1).

*Table 1: Summary of national forest inventory (NFI) plots and independent validation stands (Val) with known age; characteristics described are number of plots/stands (n), minimum (min), maximum (max), mean, and standard deviation (sd) of age and of mean height, and min, max, and mean of site index (SI), predicted SI (pSI, see next section), and area in hectare (ha).*

| | Age (years) | | | | | 95th percentile of ALS first return heights (m) [Field measured arithmetic mean height (m)] | | | | SI (pSI) | | | Area (ha) | | | |
|---|---|---|---|---|---|---|---|---|---|---|---|---|---|---|---|---|
| | n | min | max | mean | sd | Min | max | mean | sd | min | max | mean | min | max | mean | sum |
| NFI | | | | | | | | | | | | | | | | |
| spruce | 2121 | 3 | 270 | 84 | 45 | 0.3 [3.8] | 32.2 [33.8] | 14.5 [15.3] | 5.3 [4.6] | 6 (8) | 26 (23) | 13 (15) | 0.025 | 0.025 | 0.025 | 53 |
| pine | 1779 | 4 | 287 | 106 | 43 | 0.2 [5.3] | 29.3 [28.1] | 13.8 [14.2] | 4.0 [3.6] | 6 (8) | 23 (23) | 10 (15) | 0.025 | 0.025 | 0.025 | 44 |
| birch | 929 | 3 | 223 | 79 | 33 | 1.0 [4.5] | 27.1 [26.3] | 12.5 [11.8] | 4.8 [4.0] | 6 (8) | 23 (23) | 11 (15) | 0.025 | 0.025 | 0.025 | 23 |
| Vali-dation | 63 | 11 | 89 | 53 | 17 | 6.5 | 21.3 | 13.1 | 3.4 | L | M | H | 0.8 | 10.2 | 2.7 | 170 |

### 1.1.1 Independent validation data

We used stand-level observations of forest age from 63 forest stands covering a total area of 170 ha as an independent validation data set. The age of these stands was quality-assured by the local forest administration. The stands were located in central Norway in Trøndelag county (Figure 1) and their age ranged from 11 to 89 years (Table 1). The SI of the stands was reported in another system than for NFI plots and consisted of the three categories "Low", "Medium", and "High". The SI of the stands was, however, not used for predicting stand age.

62 stands were dominated by spruce, and one stand was dominated by pine according to a tree species map produced independently from this study (see Section 1.1.2). The spruce, pine, and birch proportions of the stands according to the species map ranged from 48% to 100%, from 1% to 52%, and from 1% to 39%, respectively.

### 1.1.2 Auxiliary data

Variables extracted from ALS data, a mosaic of atmospherically and topographically corrected Sentinel-2 images, and a raster of the distance to the closest coast line were used for developing models and for mapping age by applying the models. Furthermore, a site index map (pSI), and a tree species map were only used for mapping age by applying the developed models.

*Airborne laser scanning data*

ALS data were collected in several campaigns for the study area, except for high mountain ranges above the tree line. Data were collected between 2010 and 2018 with a density of 2 to 5 pulses per m$^2$, resulting in first return densities ranging between 0.5 and 36 and a mean of 8 first returns per m$^2$. A fine-resolution digital terrain model (DTM, 1 m x 1 m pixel size) was produced from the last return data by the Norwegian Mapping Authority (Kartverket 2019). The ALS point cloud was height-normalized by subtracting the DTM elevation from corresponding point cloud elevation using bi-directional interpolation. The height-normalized point cloud was used to calculate various descriptive metrics for each NFI plot based on first returns, first returns above 2 m height above ground, and last returns. The metrics included mean, variance, coefficient of variations, kurtosis and skewness of ALS return heights, 10$^{th}$, 25$^{th}$, 50$^{th}$, 75$^{th}$, 90$^{th}$, and 95$^{th}$ height percentiles, and ALS return density metrics for 10 height slices (d0 – d9). Crown coverage metrics were calculated as percentage of first returns above 2, 5 and 10 m, respectively. The DTM was resampled to 16 m x 16 m, such that the cell size corresponded approximately to the area covered by an NFI plot (250 m$^2$). From the DTM, terrain slope was computed as a raster with a cell size of 16 m x 16 m. All ALS variables, DTM elevation and slope were extracted for each NFI plot and rasters for those variables with a cell size of 16 m x 16 m were created for prediction purposes. Furthermore, the time difference between the ALS and the NFI data acquisitions was calculated for each NFI plot.

*Sentinel-2 satellite images*

The two Sentinel-2 (S2) satellites are equipped with multispectral sensors, which detect a broad electromagnetic spectrum (443 nm to 2202 nm) in 13 bands. Three of these bands (1: coastal aerosol, 9: water vapor, and 10: short-wave infrared (SWIR) cirrus) are measuring atmospheric properties and were not used in this study. S2 bottom of atmosphere (BOA) reflectance images acquired between 30 June and 31 July 2018 were mosaiced using the bands B2, B3, B4, B5, B6, B7, B8, B8A, B11, and B12. The bands in 20 m resolution were resampled to 10 m with the nearest neighbor resampling method. Cloud and shadow areas were masked out by using the classification map created by the atmospheric correction software (Sen2Cor, Louis et al. 2016). The remaining datasets were mosaiced, and color balancing was performed using the PCI Geomatics software (see

Puliti et al. (2020) for more information).

S2 variables for each NFI plot were derived by extracting the area-weighted means of the pixel values intersecting with the sample plot polygons. The normalized difference vegetation index (NDVI) was calculated as band 8 minus band 4 divided by band 8 plus band 4.

*Predicted site index*

The site index layer of the Norwegian Forest Resource Map SR16 (Astrup et al. 2019) with a 16 x 16 m pixel size was used to apply our age models. The (predicted) site index (pSI) was mapped using climate and terrain variables in a boosted regression model utilizing the SI observed on NFI plots as the response (Astrup et al. 2019). Independence of age as input to site index prediction is crucial for age modelling that utilizes site index as a predictor variable. This was the case for the pSI map which was only based on climate and terrain variables. The pSI map has a resolution of 16 m x 16 m and is freely available (Norwegian Institute of Bioeconomy Research). Weighted means of the pSI pixels intersecting with the NFI plots were calculated. These weighted means were mapped to the closest SI level (Figure 2). The RMSE and MD of the pSI were 3.9 (29.7%) and -2.3, respectively for spruce, 5.5 (54.8%) and -4.5, respectively for pine, and 5.1 (45.5%) and -3.5, respectively for birch. Clear regression towards the mean effects were visible as the lowest (6) and highest (26) SI levels never occurred in the pSI.

*Tree species map*

The tree species layer (Breidenbach et al. 2020) of the Norwegian Forest Resource Map SR16 (Astrup et al. 2019) with a 16 x 16 m pixel size was used to apply our age models for spruce, pine, and birch pixels. The tree species map was based on multitemporal S2 data and the Random Forest classifier using NFI plots as a reference. Overall accuracies of this map were 75% on plot level and 90% on stand level (Breidenbach et al. 2020).

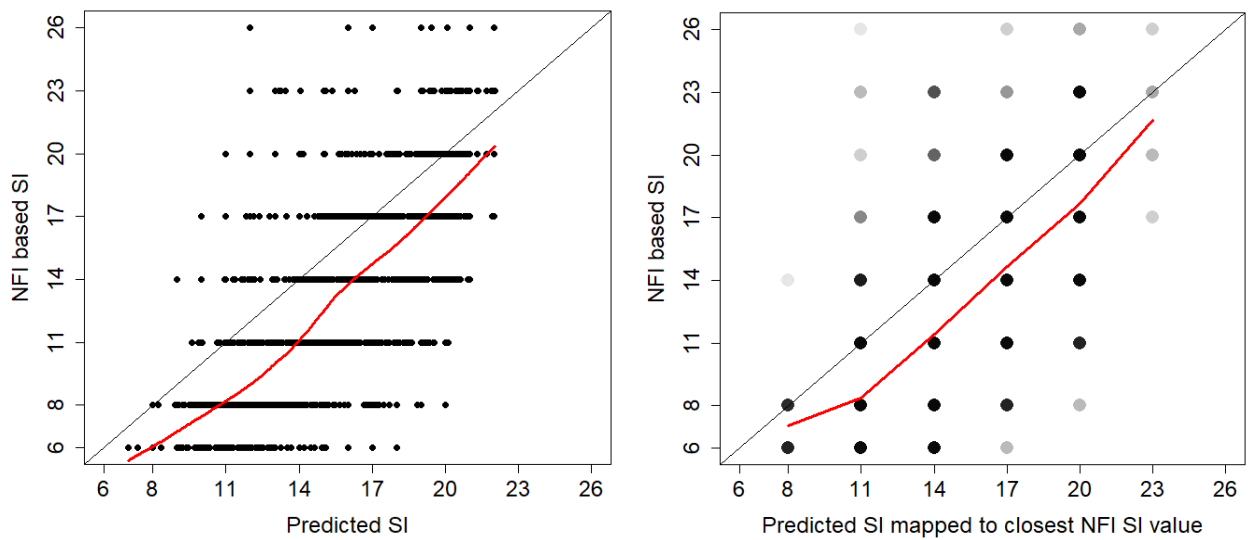

*Figure 2: NFI based site index (SI) vs. predicted SI (pSI) for spruce (plots for pine and birch were similar, and are not shown); left: SI over weighted means of pSI intersecting with NFI plots; right: SI over weighted means of pSI for each NFI plot, mapped to the closest SI level. The darker the color, the more points are over-plotted. The 1:1 line is in grey and a lowess smoothed line is in red.*

## 1.2 Methods
### 1.2.1 Age modelling

An area-based approach (Næsset 2002) was utilized to model age observed at NFI plots using remotely sensed variables as predictors. Independent linear regression models for each SI were fit with the structure

$$\boldsymbol{y} = \boldsymbol{X}\hat{\boldsymbol{\beta}} + \boldsymbol{\varepsilon}, with\ \boldsymbol{\varepsilon} \sim N(\boldsymbol{0}, \sigma^2), \tag{1}$$

where **y** = g(Age) is the n-vector of observed age with n = number of NFI plots and g as a link function, **X** = design matrix for predictor variables including an intercept, $\hat{\boldsymbol{\beta}}$ = estimated parameters, **ε** = independently and normally distributed residuals, and $\sigma^2$ = the residual variance. For each SI-specific model, we started with a model including only the 95th percentile of first ALS returns (h95_first) as a proxy for mean height. Final models were fit by forward and backward selection based on Akaike's Information Criterion (AIC) as stopping rule and further selection based on p-values (p < 0.05). We tested models with the identity (untransformed response variable), square-root, and natural logarithm as the link functions. Based on an initial analysis, the log transformation showed the best results and was chosen for further model development. We corrected for back-transformation bias by adding half the residual variance to the predictions before the back-transformation. Furthermore, we tested if adding predictors such as squared terms, and interactions between predictors improved the model.

We evaluated the models based on coefficient of determination ($R^2$), root-mean-squared-error (RMSE), and mean deviance (MD) according to

$$RMSE = \sqrt{\frac{1}{n}\sum_{i=1}^{n} w_i (y_i - \hat{y}_i)^2} \tag{2}$$

$$MD = \frac{1}{n}\sum_{i=1}^{n} w_i (y_i - \hat{y}_i) \tag{3}$$

with $w_i = 1$. Relative error statistics such as RMSE% and MD% were obtained by division by the mean of the observed values and multiplying by 100.

### 1.2.2 Validation with independent data

We evaluated our final model with the independent validation data. We mapped the stand age by applying our regression models according to the mapped tree species and pSI to the grid cells with predictor variables on a 16 m x 16 m raster. Synthetic estimates of stand age were obtained by calculating the mean predicted age for each forest stand. Finally, we compared the estimated stand age with the known age by calculating weighted versions of RMSE, RMSE%, and MD according to Equations 2 and 3, where the weights $w_i$ corresponded to the stand area proportion of the *i*th stand that sum up to 1 and n=63. All calculations were performed in R version 3.6.1 (R Core Team 2019).

## 2. Results
### 2.1 Age modelling

In the following, we will focus on results for spruce. Corresponding results for pine and birch can be found in the Appendix. The age-height relationship got stronger with increasing SI for all tree species.

More variation was observed for NFI plots in forest stands above 100 years of age, especially for SI levels below 14.

The strength of the relationship between observed and predicted forest age for the eight SI-specific models increased with increasing SI (Figure 3 for spruce; for pine and birch see Appendix Figure 8 and Figure 9). A larger variation in the predictions of the models for SI 6 and 8 was clearly visible. For spruce, the adjusted $R^2$ ranged from 0.46 to 0.96, RMSE from 2.9 to 31.2 years, RMSE% from 6.4% to 25.8%, MD from -1.0 to 2.6 years, and MD% from -2.2% to 2.2% (Table 2). Average RMSE, RMSE%, MD, and MD% for all plots classes were 21.2 years, 25.1%, 1.0 years, and 1.2%, respectively (Table 2). The results for pine and birch models were worse than for spruce, with $R^2$ ranging from 0.33 to 0.84 for pine and from 0.32 to 0.87 for birch (Appendix Table 6). For all models, the standard errors were much smaller than their corresponding parameter estimates, and most parameter estimates had p-values < 0.001. The model details for the spruce models can be found in Table 9 (for pine and birch models Table 10 and Table 11, respectively).

*Table 2: Characteristics of the fitted models for spruce. (SI: site index).*

|  | SI 6 | SI 8 | SI 11 | SI 14 | SI 17 | SI 20 | SI 23 | SI 26 | All SI |
|---|---|---|---|---|---|---|---|---|---|
|  |  |  |  | Norway spruce |  |  |  |  |  |
| $R^2$ | 0.46 | 0.60 | 0.75 | 0.79 | 0.79 | 0.84 | 0.91 | 0.96 | - |
| RMSE (Years) | 30.5 | 31.2 | 23.3 | 15.4 | 10.8 | 9.3 | 7.0 | 2.9 | 21.2 |
| RMSE (%) | 21.8 | 25.8 | 25.5 | 22.8 | 18.6 | 17.5 | 14.2 | 6.4 | 25.1 |
| MD (Years) | 2.1 | 2.6 | 1.8 | 0.5 | -0.2 | -0.4 | -0.7 | -1.0 | 1.0 |
| MD (%) | 1.5 | 2.2 | 1.9 | 0.8 | -0.3 | -0.8 | -1.5 | -2.2 | 1.2 |

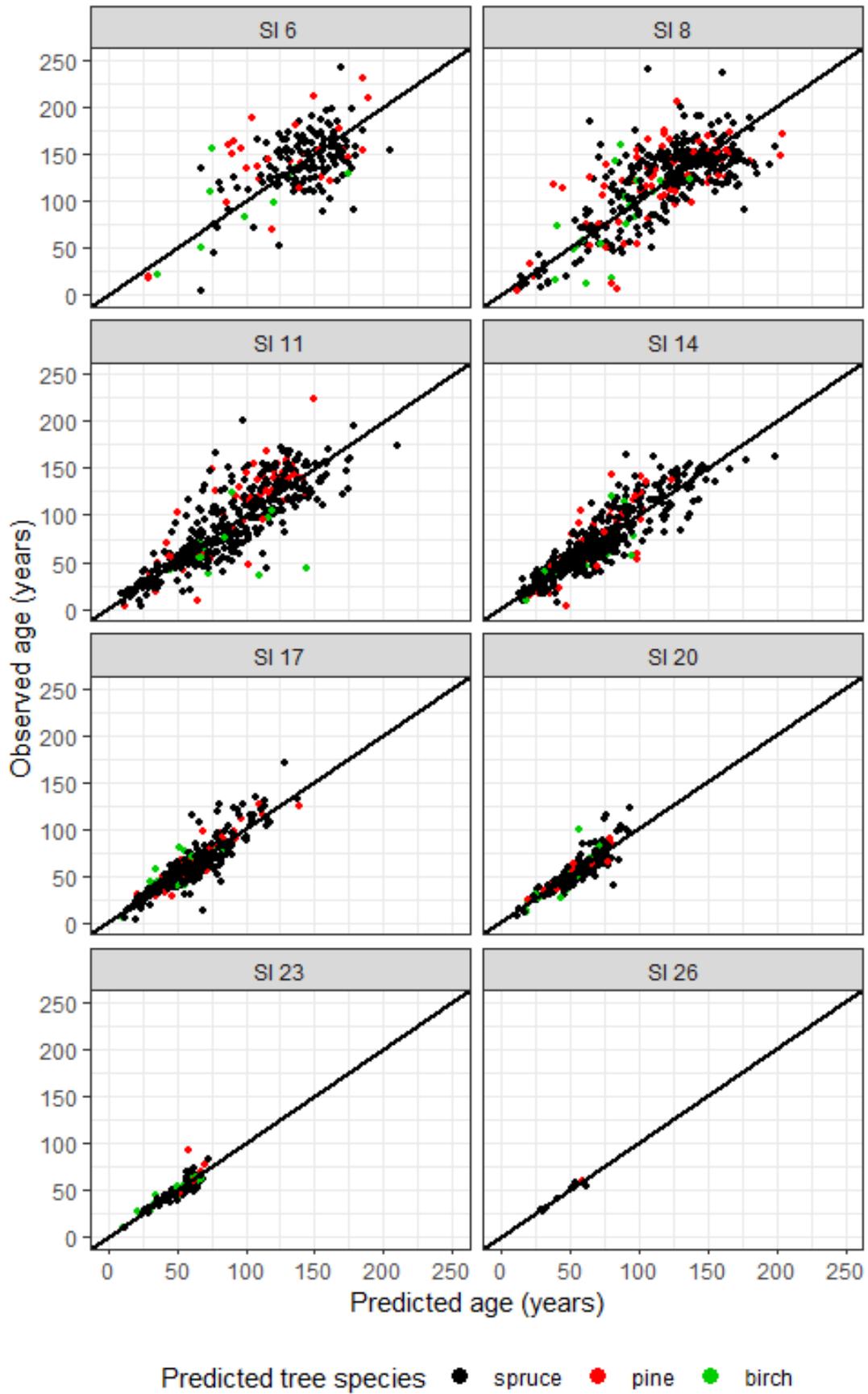

*Figure 3: Observed versus predicted age (years) in spruce forest for the eight site-index (SI) specific models with predictors from remotely sensed data; predicted tree species are presented in colour.*

All models for spruce contained the 95th percentile of the ALS first returns (h95_first) as a predictor variable. The predictor variable h95_first squared (h95_first2) was included in all models except the one for SI 26, and one of the predictor variables latitude or longitude were included in all models except the models for SI 23 and 26. The models for SI 6, 8, 14, and 17 included at least one of the spectral S2-based variables such as NDVI, S2 band 8A (s2_8A) or 11 (s2_11). The models for SI 11, 14, and 17 contained the largest number of variables with 8, 10, and 8, respectively (Appendix Table 9). Besides the already mentioned predictors, they also included crown coverage in a height of 2 m and 10 m above ground (cc2, cc10), DTM elevation, distance to the coast (distC), terrain slope (slope), and the time difference between field and ALS data acquisition (diffT). The variable h95_first was the most important predictor in all models. This was assessed by re-fitting the models with standardized predictors. The predictors were centred around their mean, and then scaled by dividing by their standard deviation. The parameter estimates of the centred and scaled version of h95_first was the largest in all models.

We assessed the model behaviour by applying the models to data where the variable h95_first ranged from 0 to the maximum observed value of this variable and all other variables were set to their mean values (Figure 4, for pine and birch see Figure 10 and Figure 11). Below a h95_first of 10 m, all models behaved similar. Above 10 m, we observed similar models for SI 6, 8, and 11, and for SI 14, 17, 20, and 23. Given h95_first, the model for SI 23 appeared more similar to the models for SI 14 and 17 than to the models for SI 20 and 26. The models, however, contain more predictors whose influence is not considered in Figure 4.

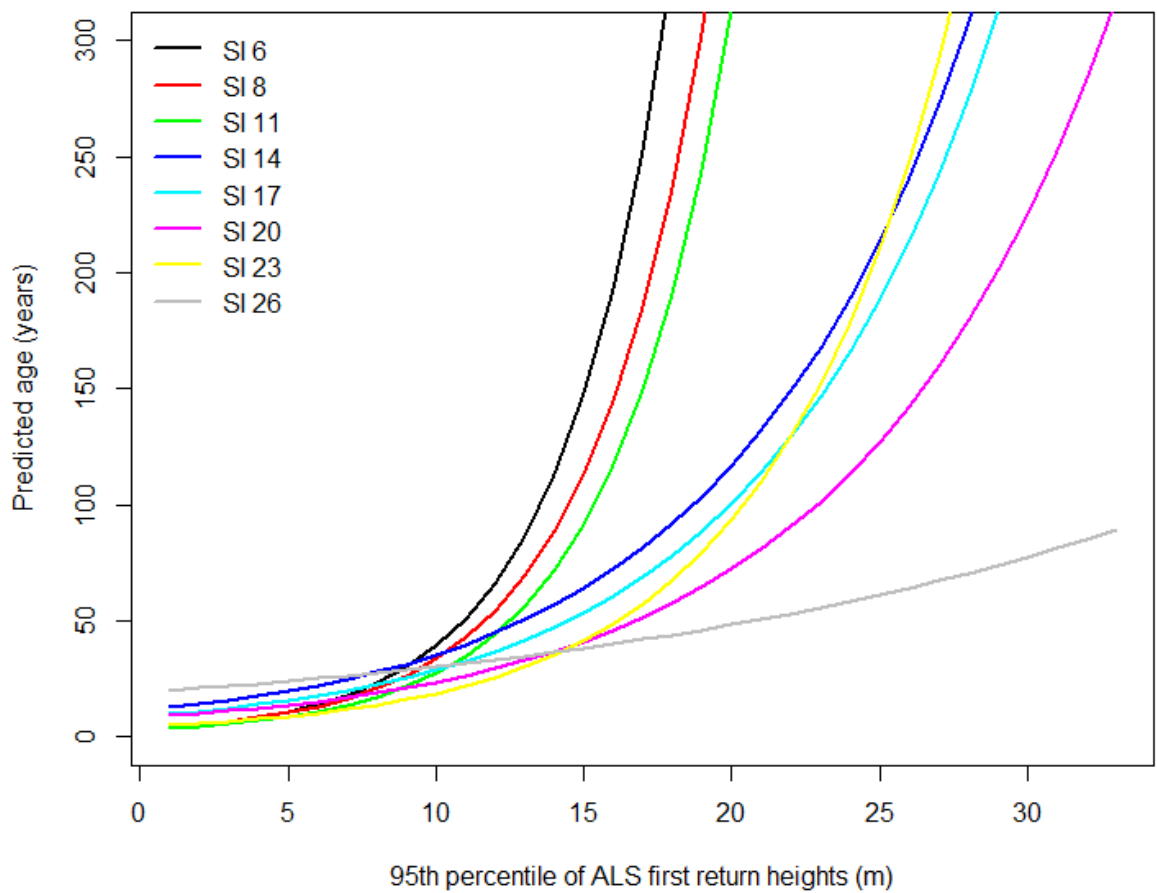

Figure 4: Predicted age (years) over the 95th percentile of first returns ALS heights (h95_first) ranging from 0 to 32, and all other predictors set to their mean values for spruce models.

We applied the SI-specific models to the NFI data using the SI map (pSI) (see Section 1.1.2) to get accuracy metrics, which are more realistic for a practical application where only pSI is available. Because pSI did not contain predictions of 6 and 26, models for these SI categories were never used. RMSE ranged from 18.6 to 56.0 years (29.4% to 53.2%), and MD from 5.4 to 37.2 years (5.3% to 31.2%). Average RMSE and MD for all pSI categories were 41.1 years (48.8%) and 20.6 years (24.5%), respectively (Table 3, for pine and birch see Table 7). Except for pSI 8, RMSE and MD decreased with increasing pSI (Table 3). Age predictions for sample plots with SI corresponding to pSI follow the 1:1 line well, whereas plots with disagreement between observed SI and pSI showed systematic lack-of-fit (Figure 5). This trend was most obvious for pSI 11 to 20. If pSI was larger than SI, the predicted age was too small (positive residual), whereas the opposite was observed if pSI was larger than SI.

Table 3: Root-mean-squared error (RMSE), RMSE%, mean deviance (MD), and MD% of the site index (SI) specific models applied using the predicted site index (pSI).

|       | pSI 8 | pSI 11 | pSI 14 | pSI 17 | pSI 20 | pSI 23 | All pSI |
|-------|-------|--------|--------|--------|--------|--------|---------|
|       |       |        | Norway spruce |  |     |        |         |
| RMSE  | 34.4  | 56.0   | 47.8   | 28.8   | 22.6   | 18.6   | 41.1    |
| RMSE% | 29.4  | 47.0   | 53.2   | 42.2   | 39.0   | 32.5   | 48.8    |
| MD    | 6.2   | 37.2   | 25.5   | 12.2   | 8.8    | 5.4    | 20.6    |
| MD%   | 5.3   | 31.2   | 28.4   | 18.0   | 15.2   | 9.4    | 24.5    |

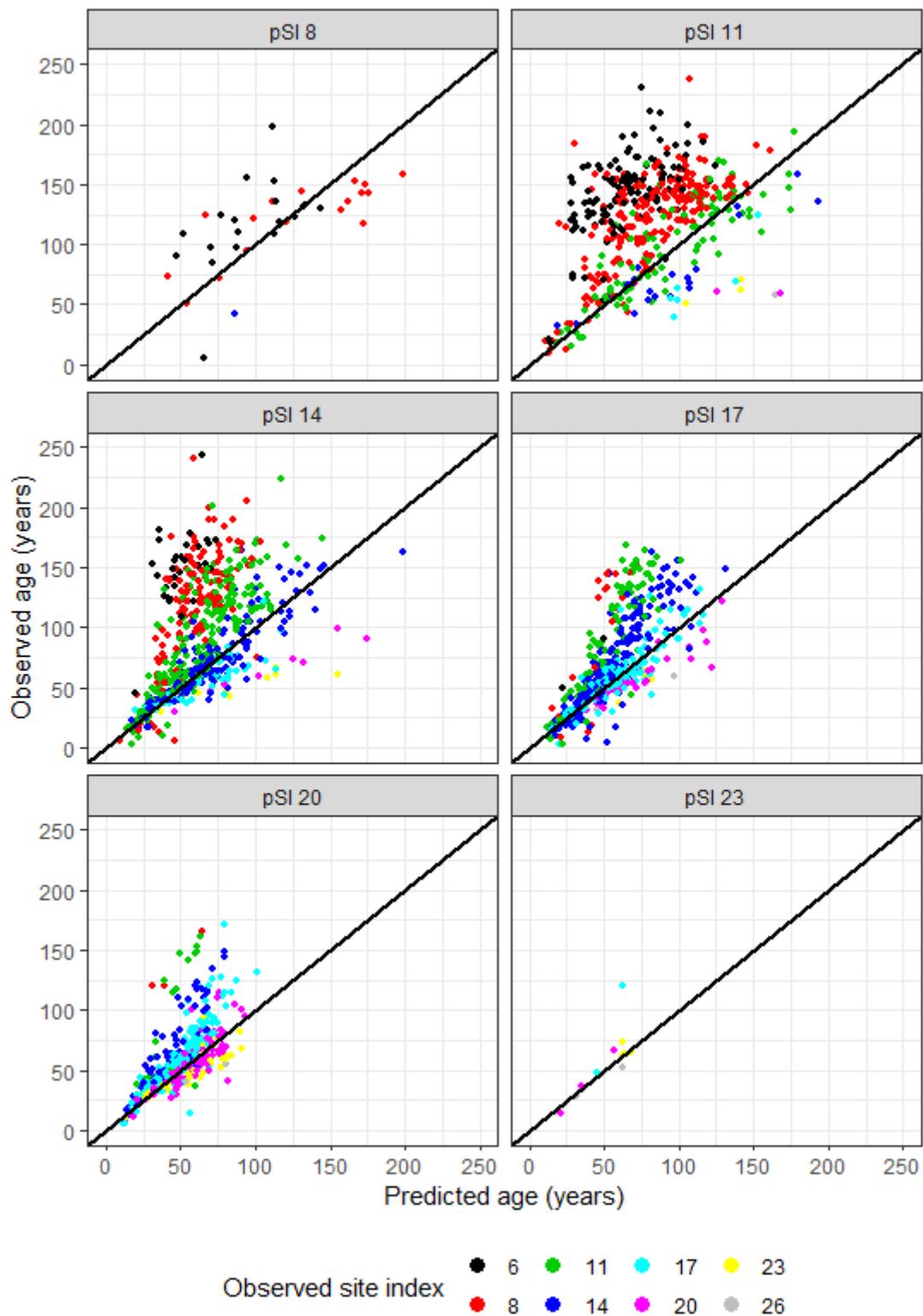

*Figure 5: Observed versus predicted age for the models applied to the six classes of predicted site index (pSI) for spruce (for pine and birch see appendix).*

In the same way as for pSI, we used mapped tree species (see Section 1.1.2) for applying the SI-specific models to the NFI data to get accuracy metrics, which are more realistic for a practical application where only a tree species map is available. To isolate the effect of predicted tree species from the pSI effect, we used observed SI for this analysis. RMSE ranged from 3 to 32 years (7% to 25%), and MD from -4 to 2 years (-4% to 2%). Average RMSE and MD for all SI categories were 21 years (25%) and 0 years, respectively (Table 4, for pine and birch see Appendix Table 8). RMSE decreased with increasing SI, however, there was no clear trend for RMSE% and MD%. Age predictions for sample plots with predicted tree species corresponding to the tree species specific model followed the 1:1 line well, whereas plots with disagreement between observed and predicted tree species often showed lack-of-fit. This trend was most obvious for pine and birch. Overall, less variability was introduced by the tree species predictions compared to pSI.

Table 4: Root-mean-squared error (RMSE), RMSE%, mean deviance (MD), and MD% of the site index (SI) specific models applied using the predicted tree species spruce.

|       | SI 6 | SI 8 | SI 11 | SI 14 | SI 17 | SI 20 | SI 23 | SI 26 | All SI |
|-------|------|------|-------|-------|-------|-------|-------|-------|--------|
|       |      |      |       | Norway spruce |       |       |       |       |        |
| RMSE  | 32.4 | 29.9 | 23.1  | 15.3  | 10.9  | 10.8  | 7.1   | 3.0   | 21.3   |
| RMSE% | 23.6 | 24.9 | 26.0  | 22.9  | 19.1  | 20.1  | 15.4  | 7.0   | 25.4   |
| MD    | -4.1 | 1.8  | 0.3   | -0.3  | -0.8  | -0.3  | -1.7  | -1.0  | -0.2   |
| MD%   | -3.0 | 1.5  | 0.4   | -0.4  | -1.5  | -0.6  | -3.7  | -2.3  | -0.2   |

## 2.2 Validation with independent field data

The estimated stand age of the validation stands resulted in area-weighted RMSE of 11.5 years (21.6%), and MD of 1.6 years (3.0%). The RMSE decreased with increasing pSI (Table 5). MD was large and negative for pSI 11, and large and positive for pSI 20. This is also reflected in the graphical representation of the results (Figure 6). For estimated pSI 11, stand ages were estimated with too large values, especially one observation with observed SI "Medium" was heavily overestimated. For estimated pSI 20, all stands with observed SI "Medium" were underestimated.

Table 5: Area-weighted root-mean-squared error (RMSE) and area weighted mean deviance (MD) separately for estimated site index (pSI) classes in 63 validation stands.

|       | pSI 11 | pSI 14 | pSI 17 | pSI 20 | All pSI |
|-------|--------|--------|--------|--------|---------|
| RMSE  | 25.7   | 12.9   | 9.0    | 8.6    | 11.5    |
| RMSE% | 48.4   | 24.4   | 16.9   | 16.2   | 21.6    |
| MD    | -23.1  | 2.8    | 1.9    | 6.6    | 1.6     |
| MD%   | -39.0  | 5.1    | 3.8    | 11.6   | 3.0     |

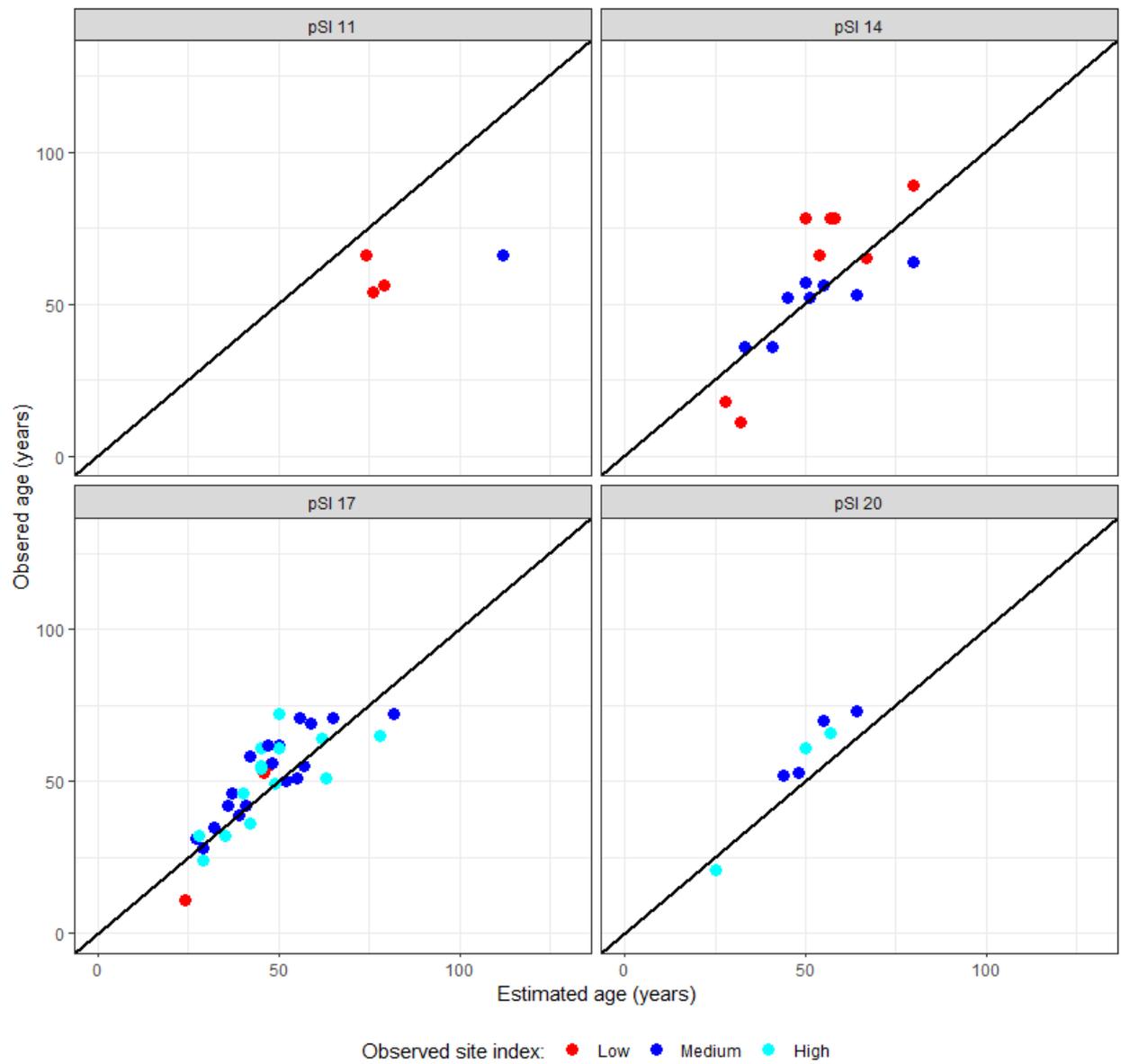

*Figure 6: Observed and estimated stand age in validation stands by estimated site indices (pSI); the observed site index (SI) on stand level was recorded as "Low", "Medium", and "High", which are presented in colour.*

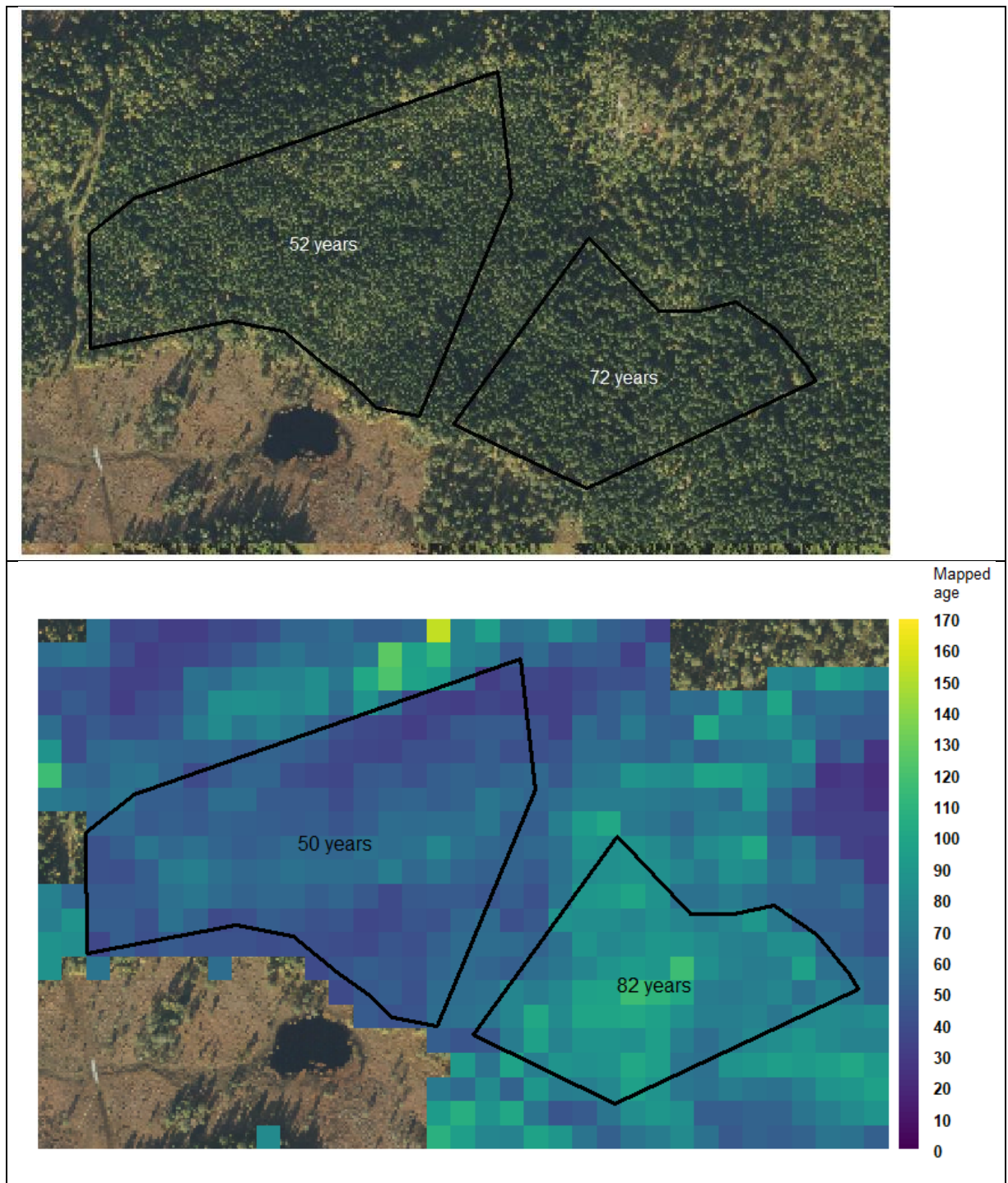

*Figure 7: Above: Outline of validation stands with known age (RGB image in background). Below: Mapped and estimated age for the same stands.*

Figure 7 provides an impression of the age map in comparison to an aerial image for two validation stands. Observed stand ages were 52 and 72 years, and site index of the stands was "Medium". Estimated stand age was 50 and 82 years and the estimated site index was 14 and 17 for the two stands, respectively.

## 3. Discussion

We mapped forest stand age using a combination of ALS and Sentinel-2 based variables over a large study area in Norway encompassing various growing conditions. We used more than 4800 NFI sample plots with stand ages ranging from 3 to over 280 years and fit SI-specific models that were validated on stand-level using independent data. We found that forest age can be predicted with relatively high accuracy especially for forest younger than 100 years and that tree height estimated from airborne laser scanning and predicted site index were the most important variables in predicting age. While the results presented in the main manuscript only describe the results for spruce, the appendix provides similar results for pine and broadleaved (birch) dominated forests. In order to apply the proposed modelling approach, site index and tree species maps are required as input variables. Errors in these maps introduced errors in the prediction of age. With improved maps of tree species and site index, our results will improve as well, especially for forest stands above 100 years of age.

We obtained the best results with SI-specific models, which showed that SI was a critical variable for describing age. We observed that the age predictions corresponded well with the reference age where observed and predicted site index agree. However, plots with observed SI smaller than pSI were underpredicted, whereas plots with observed SI larger than pSI were overpredicted. Besides uncertainties in the models, this can be explained by the fact that trees on higher SI grow faster and therefore are taller at a given age. In the opposite case, if the actual SI is smaller than the predicted SI, age at a given height is underpredicted because trees grow slower than expected. While using the mapped instead of the observed tree species hardly changed the RMSE, a 100% increase in RMSE was observed when using the mapped instead of the observed SI. This underlines the importance of an accurate site index map for using the proposed modelling approach to obtain an age map with relatively high accuracy. Nonetheless, the quality of the SI map was sufficient for obtaining reasonably good results (RMSE of 11.5 years) of stand age estimates for validation stands although errors in the SI map were noticeable also on stand level.

Other studies in boreal and temperate forests modelling forest age obtained comparable results, even though they were conducted on smaller study sites. Racine et al. (2014) used a kNN approach and reported RMSE from leave-one-out cross-validation of 8.8 years (19%) for a 66 km$^2$ study site with 158 sample plots in Canada, where mean reference age ranged from 11 to 94 years. Maltamo et al. (2009) also used a kNN approach and reported RMSE of 18.8 years (87.9%) for spruce, 23.5 years (50.7%) for pine, and 18.7 years (101.7%) for deciduous dominated sample plots in Finland, where reference age ranged from 0 to 150 years. In total they used 335 NFI sample plots in a 22,000 ha study site for modelling. A study in temperate forests was conducted by Straub & Koch (2011) who used both airborne ALS and multispectral variables to model forest stand age in a small study area covering 9.2 km$^2$, with 108 forest stands and 300 inventory sample plots in south-west Germany using linear regression. Forest age ranged from 0 to 153 years, and the forest area was composed of various deciduous and coniferous tree species. They reported an $R^2$ of 0.63, and RMSE of 19.7 years (28.8%). We obtained for more than 4800 NFI sample plots with age ranging from 3 to over 280 years RMSE between 21 and 25 years (23% to 29%). However, our study represents with almost 182,000 km$^2$ a much larger area, encompassing a wide range of growing conditions and forest structures. Our models can, therefore, be applied for practical forest management throughout the study area corresponding to the majority of the productive forests in Norway.

In the study by Maltamo et al. (2009), 69 independent validation stands with an average area of 1 ha and age ranging from 0 to 126 years were used, resulting in stand level RMSE of 18.3 years (36.3%) for spruce. In a Mediterranean to temperate climate in central Italy, Frate et al. (2015) obtained

RMSE of 16 years (30%) using 305 independent validation stands with stand age ranging from 1 to 127 years and mean of 52 years. Our results on 63 independent validation stands were comparable with RMSE of 11.5 (21.6%). The smaller errors in our study might be related to younger stands ranging from 11 to 89 years, and the better performance of our models in younger forest stands compared to older ones.

Maltamo et al. (2019, pers. communication) reported for forest stands below 100 years RMSE of 9 years. Our overall results for spruce with age up to 270 years was RMSE of 21 years. However, in an initial analysis, a model fitted with data using only NFI plots in spruce stands younger than 100 years (model not presented) resulted in smaller errors with RMSE and MD of 12.7 and 1.6 years, respectively, which fits to results by Maltamo et al. (2019, pers. communication). As results for the independent reference stands showed, our models performed well for stands below 100 years of age (Figure 6). However, 36% of the NFI plots in the productive managed forest are above 100 years, and predictions for those stands would have resulted in severe under-estimations. It was also not possible to find a satisfying model classifying older than 100 years forest from younger forest.

## 4. Conclusions

The age of spruce, pine and broadleaved (birch) dominated forest stands was mapped on a fine scale (16m x 16 m) for a large study site using variables from remotely sensed data and SI-specific models. Tree height estimated from airborne laser scanning and predicted site index were the most important variables in the models describing age. Errors decreased with increasing site index. Above 100 years of age, the model predictions had more variation and higher uncertainty. Improved site index maps would be the single most important measure to improve the age prediction.

## 5. Acknowledgements

We would like to thank the Fylkesmann in Trøndelag, represented by Arne Rannem, for providing the forest stands used for validation.

## Funding

This study was supported by the Norwegian Institute of Bioeconomy Research.

## Appendix

*Table 6: Characteristics of the fitted models for pine and birch. (SI: site index).*

|  | SI 6 | SI 8 | SI 11 | SI 14 | SI 17 | SI 20 | SI 23 | SI 26 | All SI |
|---|---|---|---|---|---|---|---|---|---|
|  |  |  |  | Scots pine |  |  |  |  |  |
| $R^2$ | 0.33 | 0.61 | 0.66 | 0.77 | 0.73 | 0.84 | 0.84 | - | - |
| RMSE (Years) | 30.0 | 27.0 | 23.5 | 17.8 | 8.1 | 8.3 | 5.5 | - | 24.5 |
| RMSE | 21.6 | 22.0 | 24.6 | 23.9 | 14.1 | 12.7 | 11.1 | - | 23.0 |
| (%) | | | | | | | | | |
| MD (Years) | 2.5 | 1.7 | 1.7 | 0.8 | -0.4 | -0.6 | -0.5 | - | 1.6 |
| MD (%) | 1.8 | 1.4 | 1.8 | 1.1 | -0.7 | -0.9 | -1.0 | - | 1.5 |
| | | | | Downy birch | | | | | |
| R² | 0.32 | 0.38 | 0.52 | 0.69 | 0.69 | 0.73 | 0.87 | - | - |
| RMSE (Years) | 26.8 | 23.7 | 23.0 | 19.3 | 16.0 | 14.8 | 4.3 | - | 22.0 |
| RMSE (%) | 28.1 | 25.7 | 28.7 | 28.8 | 26.6 | 31.3 | 11.0 | - | 27.8 |
| MD (Years) | 2.7 | 1.9 | 2.3 | 1.3 | 0.5 | 0.7 | -0.7 | - | 1.7 |
| MD (%) | 2.8 | 2.0 | 2.8 | 2.0 | 0.8 | 1.4 | -1.9 | - | 2.2 |

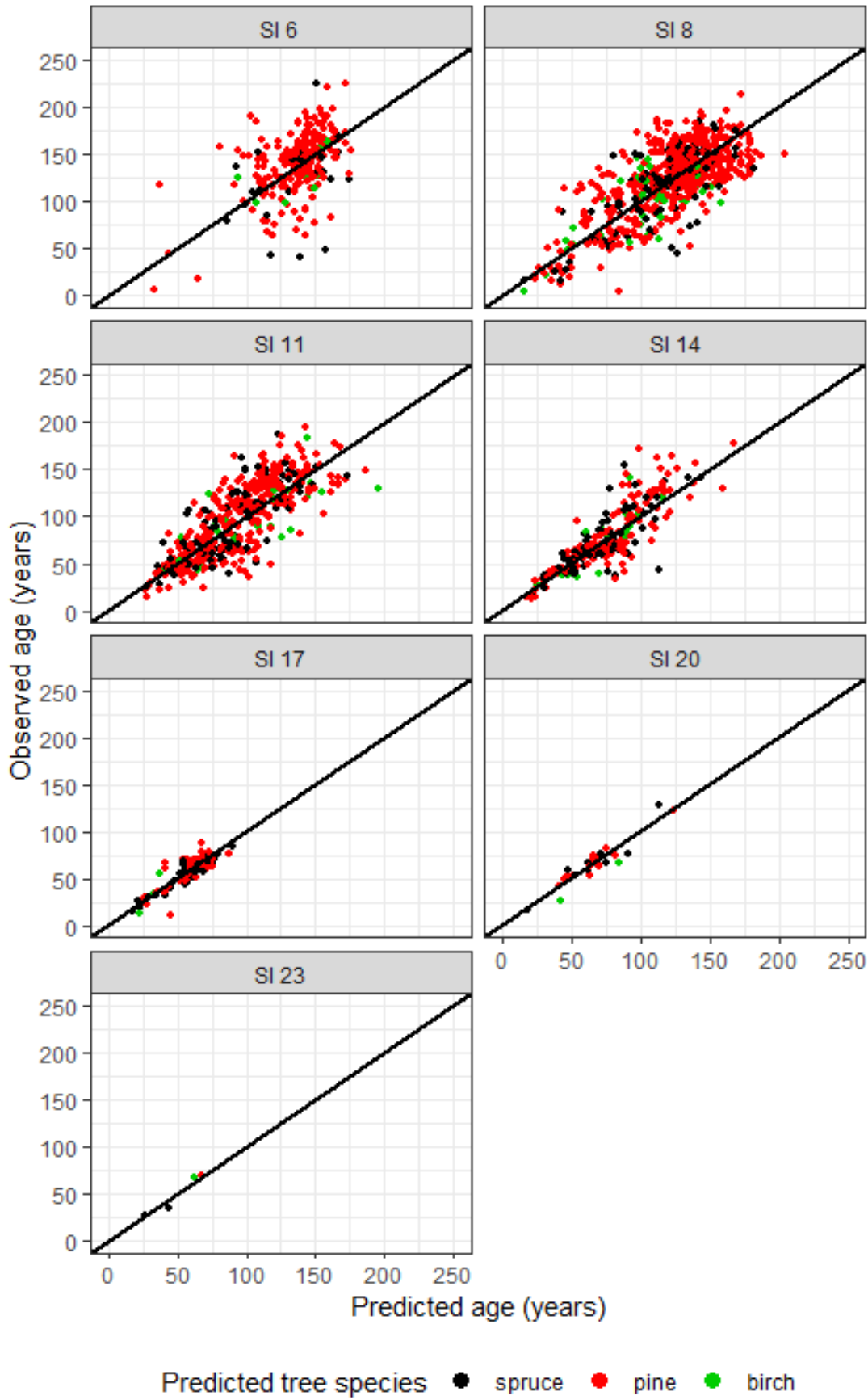

*Figure 8: Observed versus predicted age (years) for the eight site-index (SI) specific models with predictors from remotely sensed data for pine; predicted tree species in colour.*

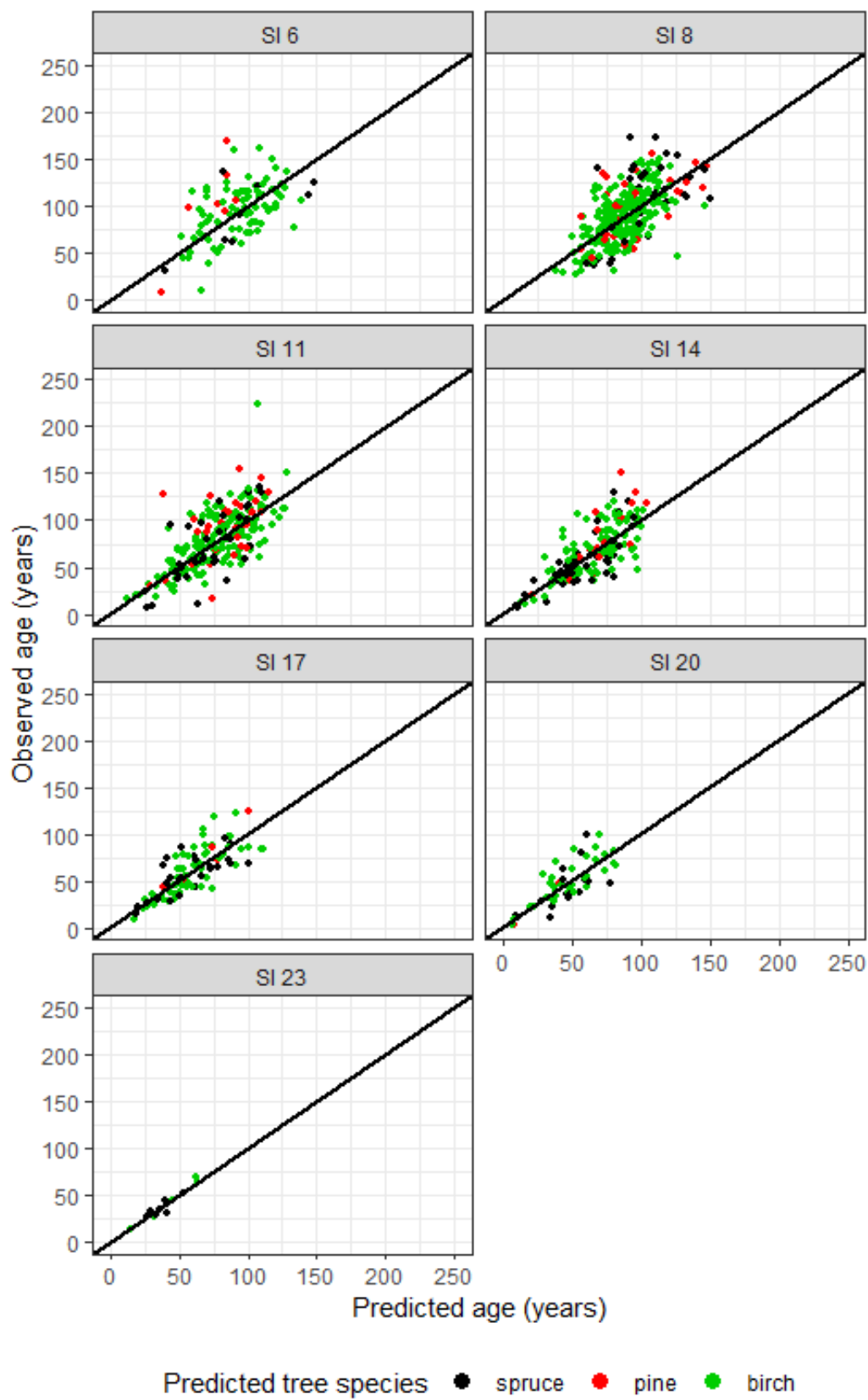

*Figure 9: Observed versus predicted age (years) for the eight site-index (SI) specific models with predictors from remotely sensed data for birch; predicted tree species in colour.*

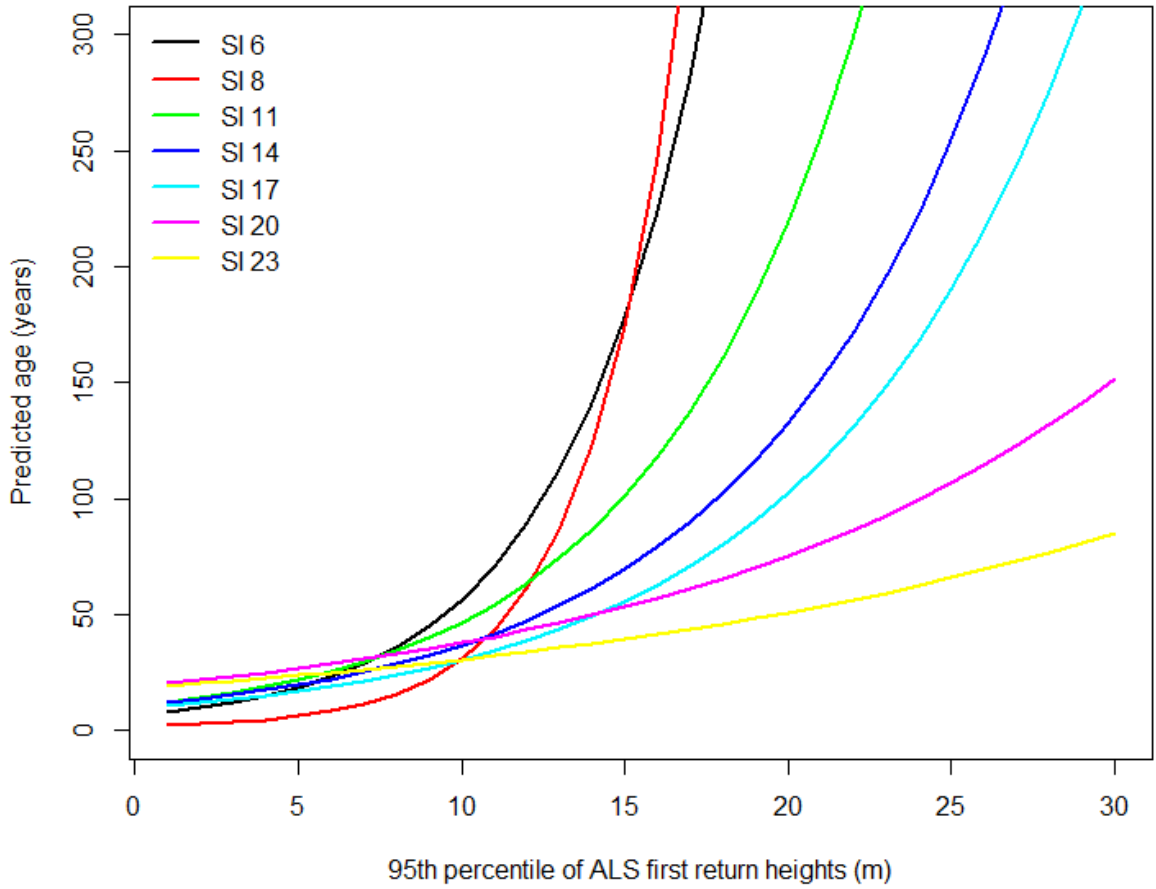

*Figure 10: Predicted age (years) over the 95th percentile of first returns ALS heights (h95_first) ranging from 0 to 32, and all other predictors set to their mean values for pine models.*

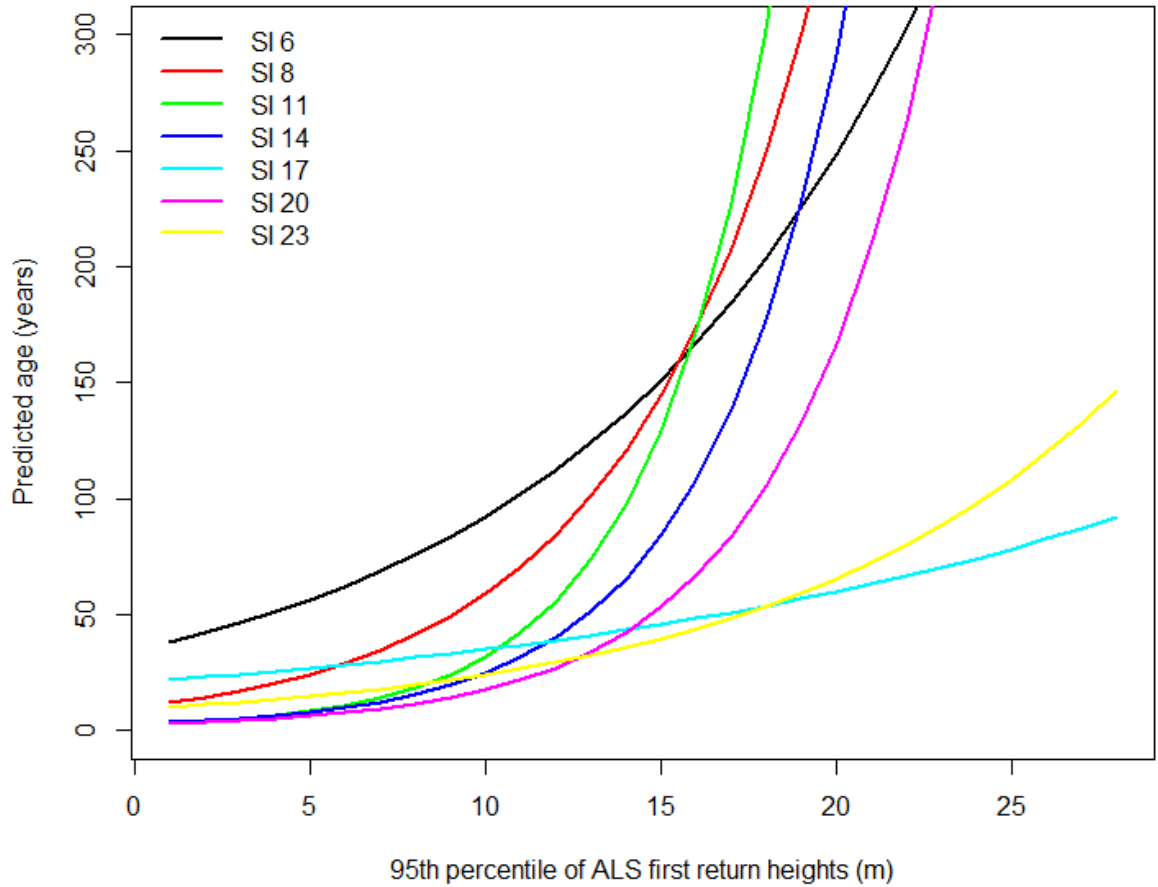

*Figure 11: Predicted age (years) over the 95$^{th}$ percentile of first returns ALS heights (h95_first) ranging from 0 to 32, and all other predictors set to their mean values for birch models.*

*Table 7: Root-mean-squared error (RMSE), RMSE%, and mean deviance (MD) of the site index (SI) specific models applied using the predicted site index (pSI) for pine and birch.*

|       | pSI 8 | pSI 11 | pSI 14 | pSI 17 | pSI 20 | pSI 23 | All pSI |
|-------|-------|--------|--------|--------|--------|--------|---------|
|       |       |        | Scots pine |    |        |        |         |
| RMSE  | 38.5  | 55.5   | 62.6   | 57.2   | 48.0   | 47.0   | 57.7    |
| RMSE% | 32.4  | 46.0   | 57.7   | 60.5   | 53.0   | 72.3   | 54.2    |
| MD    | 16.9  | 42.0   | 47.1   | 41.3   | 31.0   | 17.4   | 42.0    |
| MD%   | 14.2  | 34.8   | 43.5   | 43.7   | 34.3   | 26.8   | 39.5    |
|       |       |        | Downy birch |   |        |        |         |
| RMSE  | 25.4  | 43.9   | 40.2   | 36.2   | 36.4   | 36.7   | 37.2    |
| RMSE% | 26.7  | 51.9   | 51.3   | 49.8   | 50.8   | 60.5   | 47.1    |
| MD    | 11.4  | 24.4   | 25.0   | 21.9   | 23.0   | 23.0   | 21.7    |
| MD%   | 12.0  | 28.9   | 31.9   | 30.1   | 32.2   | 37.9   | 27.5    |

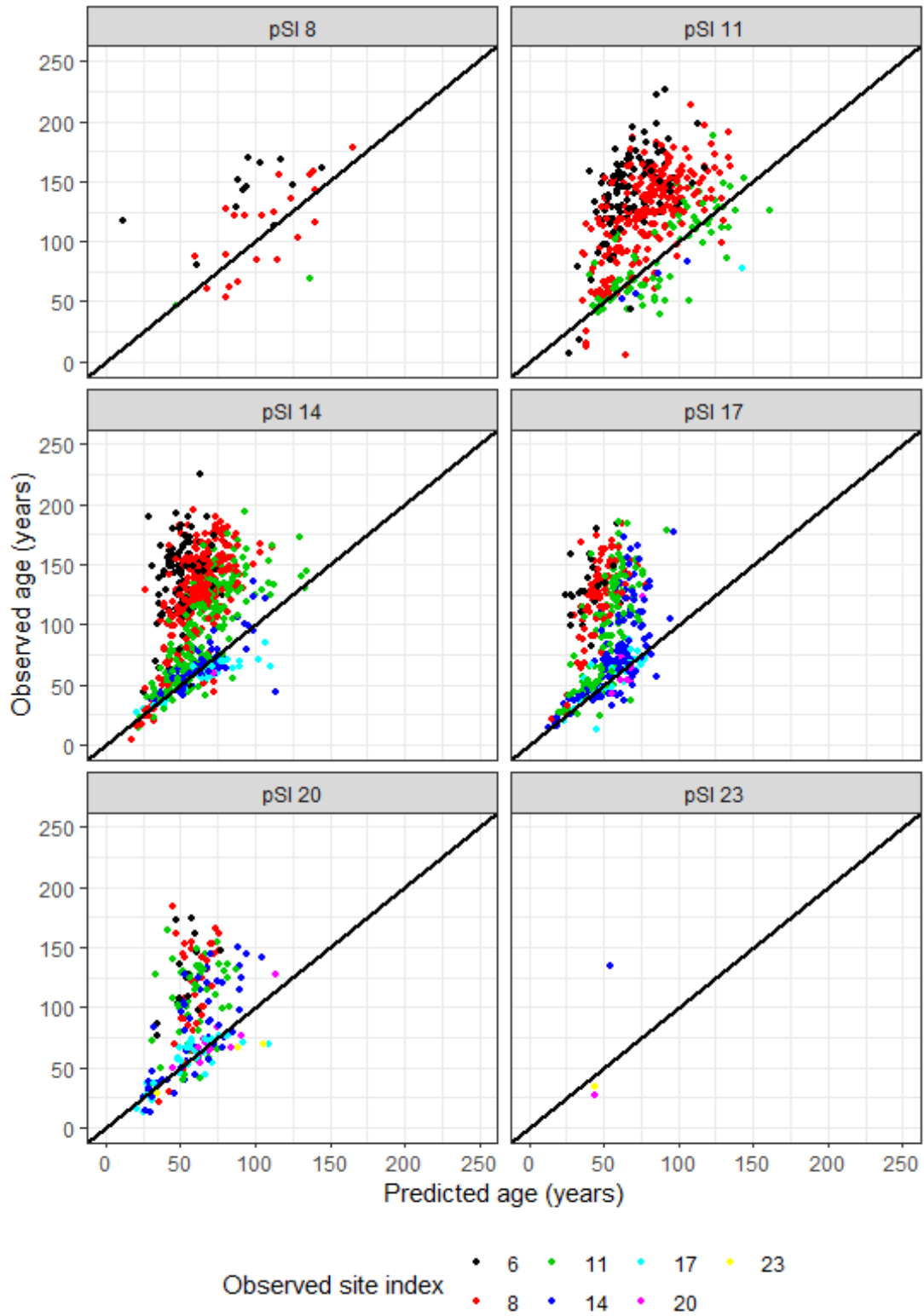

*Figure 12: Observed versus predicted age for the models applied to the six classes of predicted site index (pSI) for pine.*

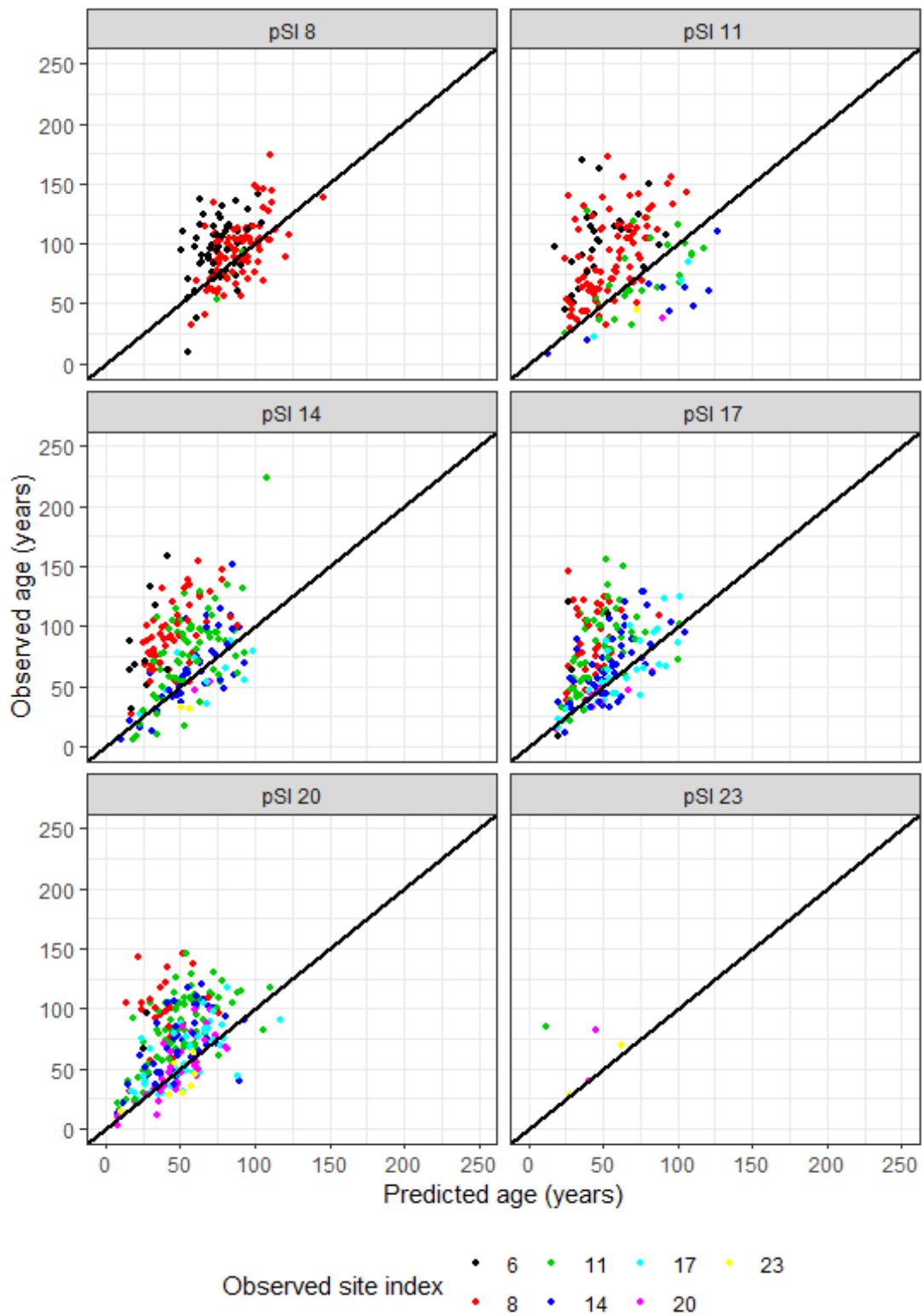

*Figure 13: Observed versus predicted age for the models applied to the six classes of predicted site index (pSI) for birch.*

Table 8: Root-mean-squared error (RMSE), RMSE%, mean deviance (MD), and MD% of the site index (SI) specific models applied using the predicted tree species pine and birch.

|  | SI 6 | SI 8 | SI 11 | SI 14 | SI 17 | SI 20 | SI 23 | All SI |
|---|---|---|---|---|---|---|---|---|
|  | | | | Scots pine | | | | |
| RMSE | 29.91 | 29.77 | 24.54 | 18.99 | 16.40 | 16.49 | 12.15 | 26.24 |
| RMSE% | 21.33 | 24.06 | 25.27 | 25.07 | 25.88 | 28.49 | 19.60 | 24.41 |
| MD | 4.98 | 3.66 | 2.75 | 0.14 | 4.50 | -5.47 | 3.41 | 3.00 |
| MD% | 3.55 | 2.96 | 2.83 | 0.18 | 7.11 | -9.46 | 5.51 | 2.78 |
|  | | | | Downy birch | | | | |
| RMSE | 27.30 | 24.27 | 23.31 | 19.35 | 15.82 | 13.74 | 14.34 | 22.34 |
| RMSE% | 27.90 | 26.90 | 28.75 | 28.43 | 27.24 | 28.32 | 32.11 | 28.27 |
| MD | 2.46 | -1.18 | 1.40 | 0.92 | 0.88 | 0.17 | 0.44 | 0.45 |
| MD% | 2.52 | -1.31 | 1.73 | 1.35 | 1.51 | 0.36 | 0.98 | 0.57 |

Table 9: Models for spruce; coefficients, their standard errors, t- and p-values for the site index (SI) specific linear regression models.

| Variable | Estimate | Std. Error | t-Value | p-value |
|---|---|---|---|---|
| **Model for SI 6** | | | | |
| *Intercept* | 2.146e+00 | 3.486e-01 | 6.157 | < 0.001 |
| *h95_first* | 2.696e-01 | 3.522e-02 | 7.654 | < 0.001 |
| *h95_first2* | -9.697e-03 | 1.762e-03 | -5.504 | < 0.001 |
| *NDVI* | 8.016e-01 | 3.168e-01 | 2.530 | 0.012 |
| *Lon* | 5.196e-02 | 1.890e-02 | 2.750 | 0.007 |
| *diffT* | -2.707e-02 | 1.230e-02 | -2.200 | 0.029 |
| Residual standard error (log scale): 0.320 | | | | |
| **Model for SI 8** | | | | |
| *Intercept* | -2.837e+00 | 8.517e-01 | -3.331 | < 0.001 |
| *h95_first* | 2.476e-01 | 2.247e-02 | 11.021 | < 0.001 |
| *h95_first2* | -6.564e-03 | 9.370e-04 | -7.005 | < 0.001 |
| *s2_8A* | -3.037e-04 | 6.148e-05 | -4.940 | < 0.001 |
| *s2_11* | 1.739e-04 | 7.440e-05 | 2.337 | 0.020 |
| *DTM* | 6.502e-04 | 1.221e-04 | 5.325 | < 0.001 |
| *distC* | -1.661e-06 | 3.834e-07 | -4.332 | < 0.001 |
| *Lat* | 9.540e-02 | 1.388e-02 | 6.876 | < 0.001 |
| Residual standard error (log scale): 0.360 | | | | |
| **Model for SI 11** | | | | |
| *Intercept* | -2.648e+00 | 6.578e-01 | -4.025 | < 0.001 |
| *h95_first* | 2.472e-01 | 1.681e-02 | 14.705 | < 0.001 |
| *h95_first2* | -4.696e-03 | 6.111e-04 | -7.684 | < 0.001 |
| *cc2* | -3.213e-01 | 1.037e-01 | -3.098 | 0.02 |
| *DTM* | -1.144e-06 | 3.143e-07 | -3.639 | < 0.001 |
| *distC* | 7.602e-02 | 1.035e-02 | 7.346 | < 0.001 |
| *Lat* | -2.828e-03 | 1.004e-03 | -2.815 | < 0.001 |
| *slope* | 2.887e-02 | 7.933e-03 | 3.640 | 0.005 |
| *diffT* | -2.648e+00 | 6.578e-01 | -4.025 | < 0.001 |
| Residual standard error (log scale): 0.321 | | | | |
| **Model for SI 14** | | | | |
| *Intercept* | -2.634e+00 | 5.356e-01 | -4.919 | < 0.001 |

| Variable | Estimate | Std. Error | t-Value | p-value |
|---|---|---|---|---|
| *h95_first* | 1.217e-01 | 1.036e-02 | 11.756 | < 0.001 |
| *h95_first2* | -1.038e-03 | 3.380e-04 | -3.072 | 0.002 |
| *NDVI* | 1.761e+00 | 3.186e-01 | 5.527 | < 0.001 |
| *s2_8A* | -3.176e-04 | 4.804e-05 | -6.610 | < 0.001 |
| *s2_11* | 4.705e-04 | 8.497e-05 | 5.538 | < 0.001 |
| *DTM* | 3.775e-04 | 9.413e-05 | 4.011 | < 0.001 |
| *distC* | -8.311e-07 | 2.508e-07 | -3.314 | 0.001 |
| *Lat* | 6.321e-02 | 7.357e-03 | 8.592 | < 0.001 |
| *slope* | -2.277e-03 | 8.371e-04 | -2.720 | 0.007 |
| *diffT* | 2.171e-02 | 5.892e-03 | 3.684 | < 0.001 |
| Residual standard error (log scale): 0.262 | | | | |
| **Model for SI 17** | | | | |
| *Intercept* | -3.354e-01 | 4.639e-01 | -0.723 | 0.470 |
| *h95_first* | 1.272e-01 | 1.267e-02 | 10.043 | < 0.001 |
| *h95_first2* | -1.264e-03 | 3.305e-04 | -3.825 | < 0.001 |
| *cc10* | -2.310e-01 | 7.248e-02 | -3.187 | 0.002 |
| *s2_8A* | -6.457e-05 | 2.597e-05 | -2.487 | 0.013 |
| *DTM* | 3.974e-04 | 9.645e-05 | 4.120 | < 0.001 |
| *distC* | -7.431e-07 | 2.340e-07 | -3.176 | 0.002 |
| *Lat* | 4.548e-02 | 7.701e-03 | 5.906 | < 0.001 |
| *diffT* | 2.790e-02 | 6.301e-03 | 4.428 | < 0.001 |
| Residual standard error (log scale): 0.215 | | | | |
| **Model for SI 20** | | | | |
| *Intercept* | 7.83e-01 | 5.22e-01 | 1.499 | 0.136 |
| *h95_first* | 1.15e-01 | 1.03e-02 | 11.105 | < 0.001 |
| *h95_first2* | -1.34e-03 | 3.00e-04 | -4.456 | < 0.001 |
| *Lat* | 2.36e-02 | 8.66e-03 | 2.721 | 0.007 |
| *slope* | -2.32e-03 | 7.51e-04 | -3.082 | 0.002 |
| *diffT* | 3.06e-02 | 7.14e-03 | 4.281 | < 0.001 |
| Residual standard error (log scale): 0.164 | | | | |
| **Model for SI 23** | | | | |
| *Intercept* | 2.02e+00 | 9.24e-02 | 21.823 | < 0.001 |
| *h95_first* | 1.65e-01 | 1.44e-02 | 11.419 | < 0.001 |
| *h95_first2* | -2.74e-03 | 3.59e-04 | -7.630 | < 0.001 |
| *cc10* | -3.92e-01 | 1.03e-01 | -3.789 | < 0.001 |
| *diffT* | 2.21e-02 | 7.66e-03 | 2.879 | 0.005 |
| Residual standard error (log scale): 0.132 | | | | |
| **Model for SI 26** | | | | |
| *Intercept* | 3.27e+00 | 2.14e-01 | 15.302 | < 0.001 |
| *h95_first* | 4.79e-02 | 3.62e-03 | 13.239 | < 0.001 |
| *cc5* | -4.66e-01 | 1.98e-01 | -2.349 | 0.057 |
| Residual standard error (log scale): 0.062 | | | | |

Table 10: Models for pine; coefficients, their standard errors, t- and p-values for the site index (SI) specific linear regression models.

| Variable | Estimate | Std. Error | t-Value | p-value |
|---|---|---|---|---|
| **Model for SI 6** | | | | |
| *Intercept* | 3.332e+00 | 1.525e-01 | 21.846 | < 0.001 |
| *h95_first* | 2.316e-01 | 2.886e-02 | 8.025 | < 0.001 |
| *h95_first2* | -8.468e-03 | 1.417e-03 | -5.977 | < 0.001 |
| *DTM* | 2.801e-04 | 7.232e-05 | 3.873 | < 0.001 |
| Residual standard error (log scale): 0.287 | | | | |
| **Model for SI 8** | | | | |
| *Intercept* | -1.139e+00 | 4.983e-01 | -2.286 | 0.023 |
| *h95_first* | 3.516e-01 | 1.957e-02 | 17.970 | < 0.001 |
| *h95_first2* | -1.034e-02 | 8.031e-04 | -12.870 | < 0.001 |
| *cc10* | 3.214e-01 | 9.059e-02 | 3.548 | < 0.001 |
| *cc2* | -5.827e-01 | 9.461e-02 | -6.159 | < 0.001 |
| *NDVI* | -4.305e-01 | 1.323e-01 | -3.255 | 0.001 |
| *DTM* | 1.731e-04 | 7.047e-05 | 2.456 | 0.014 |
| *distC* | -5.780e-07 | 2.429e-07 | -2.379 | 0.018 |
| *Lat* | 5.903e-02 | 7.780e-03 | 7.588 | < 0.001 |
| *diffT* | 1.332e-02 | 5.583e-03 | 2.387 | 0.017 |
| Residual standard error (log scale): 0.287 | | | | |
| **Model for SI 11** | | | | |
| *Intercept* | -2.296e-01 | 5.697e-01 | -0.403 | 0.687 |
| *h95_first* | 1.566e-01 | 2.393e-02 | 6.545 | < 0.001 |
| *h95_first2* | -2.144e-03 | 7.813e-04 | -2.744 | 0.006 |
| *cc10* | 3.698e-01 | 9.589e-02 | 3.857 | < 0.001 |
| *cc2* | -6.925e-01 | 9.942e-02 | -6.966 | < 0.001 |
| *s2_11* | 2.182e-04 | 5.943e-05 | 3.671 | < 0.001 |
| *DTM* | 2.202e-04 | 8.549e-05 | 2.575 | 0.010 |
| *distC* | -1.298e-06 | 2.751e-07 | -4.719 | < 0.001 |
| *Lat* | 4.614e-02 | 9.326e-03 | 4.947 | < 0.001 |
| *slope* | 1.949e-03 | 7.736e-04 | 2.519 | 0.012 |
| *diffT* | 1.492e-02 | 6.189e-03 | 2.411 | 0.016 |
| Residual standard error (log scale): 0.270 | | | | |
| **Model for SI 14** | | | | |
| *Intercept* | -3.026e+00 | 1.090e+00 | -2.776 | 0.006 |
| *h95_first* | 1.304e-01 | 1.609e-02 | 8.105 | < 0.001 |
| *h95_first2* | -9.890e-04 | 5.068e-04 | -1.951 | 0.052 |
| *cc2* | -2.876e-01 | 1.051e-01 | -2.737 | 0.007 |
| *NDVI* | 1.561e+00 | 4.365e-01 | 3.577 | < 0.001 |
| *s2_8A* | -2.633e-04 | 8.332e-05 | -3.161 | 0.002 |
| *s2_11* | 6.118e-04 | 1.413e-04 | 4.329 | < 0.001 |
| *distC* | -1.056e-06 | 3.070e-07 | -3.439 | 0.001 |
| *Lat* | 7.074e-02 | 1.635e-02 | 4.328 | < 0.001 |
| *diffT* | 2.126e-02 | 8.076e-03 | 2.632 | 0.009 |
| Residual standard error (log scale): 0.230 | | | | |
| **Model for SI 17** | | | | |
| *Intercept* | -4.554e+00 | 1.869e+00 | -2.437 | 0.017 |
| *h95_first* | 1.242e-01 | 3.267e-02 | 3.803 | < 0.001 |
| *h95_first2* | -2.344e-03 | 9.885e-04 | -2.372 | 0.020 |
| *cc10* | 2.865e-01 | 1.414e-01 | 2.026 | 0.046 |
| *NDVI* | 1.561e+00 | 6.292e-01 | 2.481 | 0.015 |
| *s2_8A* | -2.120e-04 | 1.036e-04 | -2.046 | 0.044 |
| *s2_11* | 4.390e-04 | 1.743e-04 | 2.518 | 0.014 |
| *Lat* | 9.259e-02 | 2.656e-02 | 3.487 | 0.001 |
| *diffT* | 3.027e-02 | 1.040e-02 | 2.911 | 0.005 |
| Residual standard error (log scale): 0.208 | | | | |

| Variable | Estimate | Std. Error | t-Value | p-value |
|---|---|---|---|---|
| **Model for SI 20** | | | | |
| *Intercept* | -2.037e+00 | 2.176e+00 | -0.936 | 0.361 |
| *h95_first* | 7.050e-02 | 7.744e-03 | 9.104 | < 0.001 |
| *cc2* | -4.554e-01 | 1.878e-01 | -2.425 | 0.025 |
| *Lon* | -1.070e-01 | 2.583e-02 | -4.142 | 0.001 |
| *Lat* | 1.055e-01 | 3.637e-02 | 2.901 | 0.009 |
| *slope* | -8.509e-03 | 3.342e-03 | -2.546 | 0.020 |
| Residual standard error (log scale): 0.163 | | | | |
| **Model for SI 23** | | | | |
| *Intercept* | 2.865e+00 | 2.552e-01 | 11.224 | 0.008 |
| *h95_first* | 5.231e-02 | 1.294e-02 | 4.042 | 0.056 |
| Residual standard error (log scale): 0.187 | | | | |

*Table 11: Models for birch; coefficients, their standard errors, t- and p-values for the site index (SI) specific linear regression models.*

| Variable | Estimate | Std. Error | t-Value | p-value |
|---|---|---|---|---|
| **Model for SI 6** | | | | |
| *Intercept* | -1.257e+00 | 1.903e+00 | -0.660 | 0.511 |
| *h95_first* | 9.995e-02 | 1.700e-02 | 5.879 | < 0.001 |
| *DTM* | 6.262e-04 | 1.817e-04 | 3.447 | 0.001 |
| *Lat* | 7.417e-02 | 3.105e-02 | 2.389 | 0.019 |
| Residual standard error (log scale): 0.378 | | | | |
| **Model for SI 8** | | | | |
| *Intercept* | -2.538E-01 | 6.402E-01 | -0.396 | 0.692 |
| *h95_first* | 1.819E-01 | 3.669E-02 | 4.958 | < 0.001 |
| *h95_first2* | -3.691E-03 | 1.651E-03 | -2.236 | 0.026 |
| *cc2* | -6.913E-01 | 1.098E-01 | -6.295 | < 0.001 |
| *s2_8A* | -8.220E-05 | 2.570E-05 | -3.199 | 0.002 |
| *DTM* | 3.849E-04 | 1.117E-04 | 3.445 | 0.001 |
| *distC* | -1.223E-06 | 4.300E-07 | -2.844 | 0.005 |
| *Lat* | 6.285E-02 | 1.031E-02 | 6.095 | < 0.001 |
| Residual standard error (log scale): 0.280 | | | | |
| **Model for SI 11** | | | | |
| *Intercept* | -6.228e-01 | 8.749e-01 | -0.712 | 0.477 |
| *h95_first* | 2.850e-01 | 3.334e-02 | 8.550 | < 0.001 |
| *h95_first2* | -6.689e-03 | 1.179e-03 | -5.673 | < 0.001 |
| *cc2* | -5.520e-01 | 1.497e-01 | -3.686 | < 0.001 |
| *DTM* | 3.754e-04 | 1.540e-04 | 2.437 | 0.016 |
| *distC* | -1.905e-06 | 5.907e-07 | -3.224 | 0.001 |
| *Lat* | 4.548e-02 | 1.365e-02 | 3.333 | 0.001 |
| Residual standard error (log scale): 0.353 | | | | |
| **Model for SI 14** | | | | |
| *Intercept* | -9.998e-01 | 9.065e-01 | -1.103 | 0.272 |
| *h95_first* | 2.500e-01 | 2.893e-02 | 8.641 | < 0.001 |
| *h95_first2* | -5.157e-03 | 9.403e-04 | -5.484 | < 0.001 |
| *cc2* | -3.962e-01 | 1.636e-01 | -2.421 | 0.017 |
| *distC* | -2.029e-06 | 5.904e-07 | -3.436 | 0.001 |
| *Lat* | 4.953e-02 | 1.443e-02 | 3.433 | 0.001 |
| Residual standard error (log scale): 0.302 | | | | |
| **Model for SI 17** | | | | |
| *Intercept* | -1.340E+00 | 1.154E+00 | -1.161 | 0.249 |
| *h95_first* | 5.467E-02 | 9.514E-03 | 5.746 | 0.039 |
| *cc10* | 3.418E-01 | 1.436E-01 | 2.380 | 0.020 |
| *s2_8A* | -1.323E-04 | 4.262E-05 | -3.104 | 0.003 |
| *DTM* | -4.444E-04 | 2.214E-04 | -2.007 | 0.048 |
| *Lon* | -9.251E-02 | 1.899E-02 | -4.872 | < 0.001 |
| *Lat* | 9.168E-02 | 2.011E-02 | 4.559 | < 0.001 |
| Residual standard error (log scale): 0.270 | | | | |
| **Model for SI 20** | | | | |
| *Intercept* | 1.646e+00 | 3.983e-01 | 4.133 | < 0.001 |
| *h95_first* | 2.305e-01 | 3.451e-02 | 6.678 | < 0.001 |
| *h95_first2* | -4.256e-03 | 1.121e-03 | -3.796 | < 0.001 |
| *Lon* | -7.624e-02 | 3.573e-02 | -2.133 | 0.038 |
| *diffT* | 7.409e-02 | 3.402e-02 | 2.178 | 0.034 |
| Residual standard error (log scale): 0.362 | | | | |
| **Model for SI 23** | | | | |
| *Intercept* | 8.081e+00 | 1.760e+00 | 4.591 | 0.004 |
| *h95_first* | 1.022e-01 | 1.609e-02 | 6.352 | 0.001 |
| *NDVI* | -7.766e+00 | 2.091e+00 | -3.713 | 0.010 |
| *s2_8A* | 6.981e-04 | 2.478e-04 | 2.818 | 0.030 |
| *s2_11* | -1.603e-03 | 5.395e-04 | -2.970 | 0.025 |
| *DTM* | 2.346e-03 | 7.950e-04 | 2.951 | 0.026 |
| Residual standard error (log scale): 0.159 | | | | |